\documentclass[%
 reprint,
 amsmath,amssymb,
 aps,
]{revtex4-2}
\usepackage{amsmath}
\usepackage{graphicx}
\usepackage{dcolumn}
\usepackage{bm}

\usepackage{amsmath}
\begin{document}

\preprint{APS/123-QED}

\title{A High-Q Vertical Light Emission From Active Parity-Time Symmetric Gratings}

	\author{Tahere Hemati}	
\affiliation{School of Electrical and Computer Engineering, University of Oklahoma, Norman OK 73019, United States}
	\author{Yi Zou}
	\email{zouyi@shanghaitech.edu.cn}
	\affiliation{School of Information Science and Technology, ShanghaiTech University, Shanghai 201201, China}
	\author{Binbin Weng}
	\email{binbinweng@ou.edu}
	\affiliation{School of Electrical and Computer Engineering, University of Oklahoma, Norman OK 73019, United States}




\date{\today}

\begin{abstract}

This work presents a theoretical investigation of an active photonic grating of the parity-time (PT) symmetric architecture. The analytical study of the free-space mode propagation from the grating structure indicates the unique bifurcation property due to the PT-symmetry modulation. It is shown that both the gain/loss contrast and the lattice constant parameters are critical factors to modulate the active photonic system in between the PT-symmetry to the symmetry-broken phases. Furthermore, numerical simulations via the Rigorous Coupled-Wave Analysis (RCWA) method discover the existence of a unique Spectral Singularity (SS) phenomenon in this PT grating structure, which corresponds to a non-trivial single-mode and near-zero bandwidth photonic resonant emission. Also, the guiding procedure for fulfilling SS modes is found to be related to the unique formation of the scattering matrix applied in the PT-symmetric photonic gratings. This theoretical work takes a fresh look into the active PT-symmetric photonic gratings focusing on the discovery of new free-space emission modes rather than the commonly studied unidirectional properties, which could contribute to the development of novel low-threshold and super-coherent surface-emitting devices.

\end{abstract}

\maketitle


\section{\label{sec:level1}Introduction}
Photonic gratings as the first optical components operating at subwavelength region can be considered the aspiration of modern nano-photonics concepts such as photonic crystals and metamaterials \cite{bonod2016diffraction}. The main purpose of all these optical components is to control the dynamic of light, especially on a nanometer scale \cite{loewen2018diffraction}. Today, the applications of gratings (e.g., diffraction) are expanded from telecommunications \cite{milner2005high} and astronomy \cite{barden1998volume} to chemistry \cite{ye2010diffraction} and biosensing \cite{wark2007nanoparticle}. The recent progress in nano-fabrication technologies enabled us to create highly precise grating structures and revealed the potential of these optical components in practice. Thus, in the recent decade, special attention has been attracted to this field. 

The unique properties of photonic gratings are associated with the spatial modulation of the Refractive Index (RI). By optimizing grating parameters such as the period, thickness and fill factor, we can modulate the amplitude and phase of the incident wave and design the gratings for various purposes, including optical filters \cite{nishi1985broad}, reflectors \cite{almuneau2010high, hemati2018theoretical}, absorbers \cite{shi2015compact}, resonators \cite{kaminow1971poly, hemati2019theoretical}, and so on.

More recently, by drawing the fundamental concepts in quantum physics (i.e., Parity-Time (PT) symmetry) to the realm of optics and photonics, novel applications, such as unidirectional emission \cite{feng2013experimental} and asymmetric diffraction \cite{yang2019experimental}, appeared from periodical structures, including photonic gratings.
The simplest definition of a PT-symmetric system is a non-Hermitian system whose Hamiltonian remains invariant under spatial reflection ($p$  $\xrightarrow{}$ $-p$, $x$  $\xrightarrow{}$ $-x$) and time-reversal  ($p$  $\xrightarrow{}$ $-p$, $x$  $\xrightarrow{}$ $x$, $i$  $\xrightarrow{}$ $-i$) operators, where $p$, $x$, and $i$ are momentum, location, and imaginary unite, respectively \cite{bender1998real}. It is indicated that to satisfy this condition, the related potential should be  $\hat{V}(x)=\hat{V}^*(-x)$. The equivalent condition for a optical system is ${n}(x)={n}^*(-x)$ \cite{ruter2010observation}. 

In practice, two identical coupled waveguides, one showing loss and the other one indicating the same amount of gain can be the simplest example of an optical PT-symmetric system \cite{ruter2010observation}. In this system, by increasing the gain(loss) coupling contrast between these two waveguides, the system transits from a PT-symmetric regime where all eigenvalues are real into a PT-broken regime where the eigenvalues are complex.

There is a transient point between these two regimes named exceptional point, where both eigenvalues and eigenstates coalesce \cite{ozdemir2019parity}. The uniqueness of a PT-symmetric system lies in indicating real eigenvalues, despite the non-Hermiticity and non-orthogonality of its eigenstates \cite{feng2017non}, which lead to extraordinary phenomena. Indeed, in PT-symmetric systems, we take advantage of the beneficial role of loss, while in trivial systems, loss always has a destructive effect, and it should be eliminated.

Scholars gradually found that using more complicated optical platforms such as periodical structures to realize the PT-symmetry concept can introduce new non-trivial outcomes, including loss-induced transparency \cite{guo2009observation}, non-reciprocal light propagation \cite{ramezani2010unidirectional}, unidirectional invisibility \cite{lin2011unidirectional, feng2013experimental}, and directional emission \cite{kim2014partially}. Among different kinds of periodical structures, only a few works have focused on grating structures as a  platform for exploring PT-symmetric effects \cite{zhu2016asymmetric, gao2018intrinsic, kulishov2012free, kulishov2015analysiss}, and the accomplished works mostly investigate asymmetric diffraction \cite{zhu2016asymmetric} and unidirectional properties \cite{kulishov2015analysis}. 

Herein, we instead investigate an ``active'' type of the PT-symmetric gratings and study the possible new non-trivial phenomena besides the asymmetric diffraction effect. This research will theoretically and numerically study the emission properties of diffracted modes from an active grating by Helmholtz equation, driving the related Hamilton and indicating symmetry, symmetry-broken and exceptional point regions. Moreover, the properties of resonating modes induced by PT-symmetric diffraction grating will be discussed in detail.

\section{\label{sec:level2}Theory and analytical solutions}

As just mentioned, the guiding procedure in gratings lies in the spatial RI modulation. On the other hand, to realize the PT-symmetry, the RI modulation needs to satisfy  ${n}(x)={n}^*(-x)$. Therefore, the real part of RI modulation has to be a symmetric function, and its imaginary part should be an asymmetric function. Thus the RI modulation can be defined as:
\begin{equation}
    n(x)=n_0+n\text{cos}(2\pi x/a)+i\gamma \text{sin}(2\pi x/a)
    \label{equation 1}
\end{equation}
where $a$ is the period, $n_0$ is the effective refractive index, and
$n$ and $\gamma$ are the real and imaginary parts of the refractive index modulation, respectively. 

\begin{figure}[ht]
  \centering
   {\includegraphics[scale=0.55]{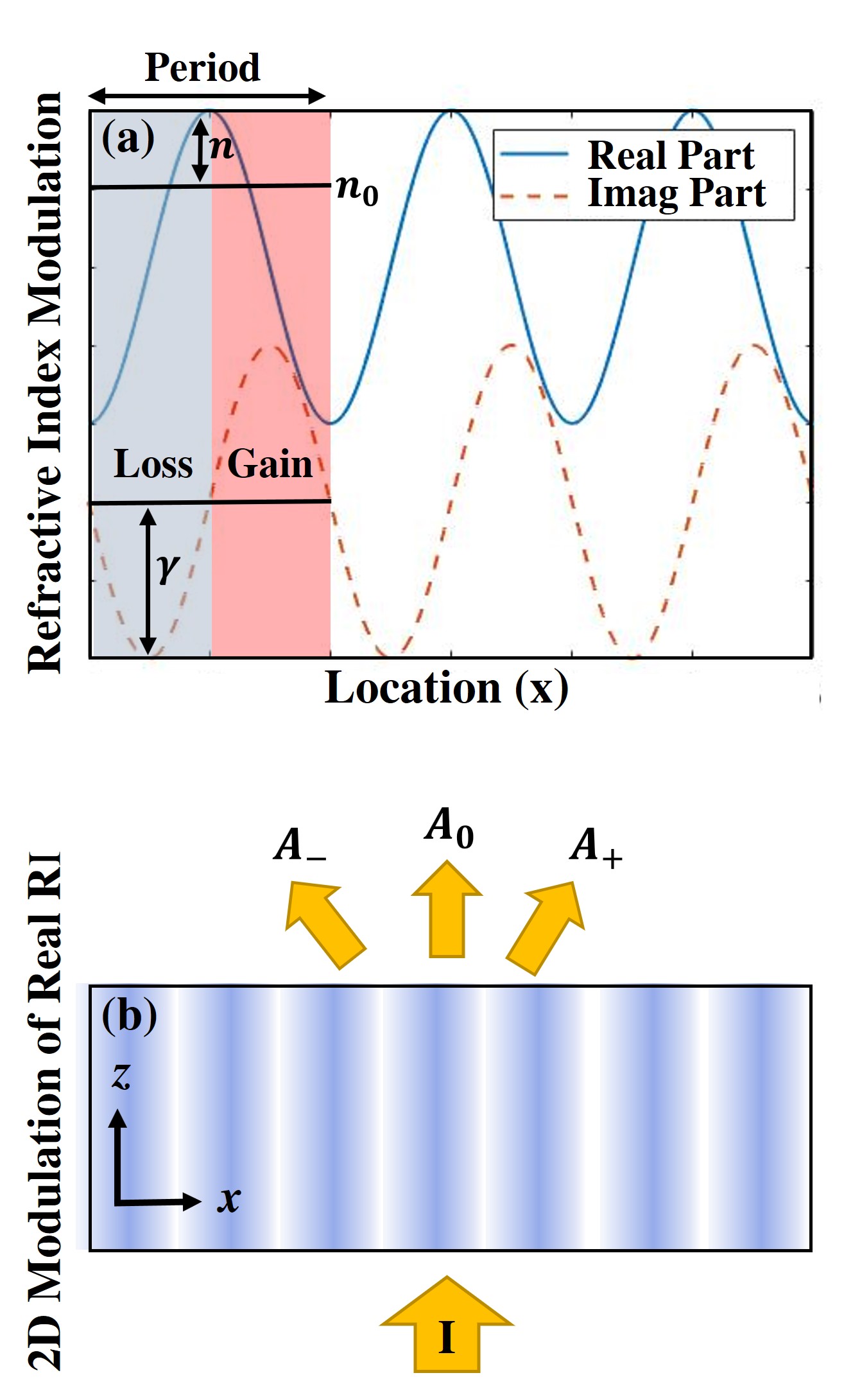}}
  \caption{(a) The real and imaginary parts of a PT-symmetric RI modulation in the x-direction. (b) The real RI modulation in x- and z-direction. }
  \label{fig:figure 1}
\end{figure}

Fig. \ref{fig:figure 1}(a) shows the real (solid blue line) and imaginary (dashed red line) parts of the PT-symmetric refractive index distribution. The positive part of the imaginary RI indicates gain, and the negative part shows the induced loss. Fig. \ref{fig:figure 1}(b) displays the real part of the RI modulation in two dimensions where the RI solely changes in the x-direction, and the incident light is perpendicular to the periodicity. $A_-$, $A_0$, and $A_+$ show the amplitude of the $-1$, $0$, and $+1$ orders of diffracted grating modes, respectively.

Here, we study the wave propagation of an incident light entering along the z-axis where the wave $E(x, z)$ obeys the Helmholtz equation \cite{gaylord1985analysis, shui2018asymmetric}:

\begin{equation}
\nabla^2E+k_0^2n^2(x)E=0
\label{equation 2}
\end{equation}
in which $k_0=2\pi/\lambda$. Since the incident light is normal to the periodicity, both x and z components are effective. The general solution for $E(x,z)$ can be derived from Bloch's wave equation. 

\begin{equation}
    E(x, z)=u(x)e^{i\vec{\beta}.\vec{r}}
    \label{equation 3}
\end{equation}
where $u(x)$ is a periodic function in the x-direction
\begin{equation}
   u(x)=\sum_{m=-\infty}^{\infty} A_m e^{-i2\pi mx/a} 
    \label{equation 4}
\end{equation}
 and in a 2D plane
 
 \begin{equation}
    e^{i\vec{\beta}.\vec{r}}=e^{i\beta_xx}e^{i\beta_zz}
    \label{equation 5}
\end{equation}
By substituting Eq.~(\ref{equation 4}) and Eq.~(\ref{equation 5}) into Eq.~(\ref{equation 3}):

\begin{equation}
   E(x,z)= \sum_{m=-\infty}^{\infty}A_m e^{i\beta_zz} e^{i(\beta_x-2\pi m/a)x}
    \label{equation 6}
\end{equation}

Considering the normal light incident, the wavenumber in  x-direction equals $2\pi m/a$, and consequently, the dispersion relation leads to:

\begin{equation}
   \beta_z=\sqrt{k_{0}^2n_0^2-({2\pi m/a)}^2}
    \label{equation 7}
\end{equation}
If we consider the grating in a diffracted regime where $a>>\lambda$, $\beta_z$ will be reduced to $\beta_z=k_0n_0$. Thus, Eq.~(\ref{equation 6}) can be rephrased as:
 \begin{equation} \label{equation 8}
E(x, z)=\sum_{m=-\infty}^{\infty} A_m e^{-i[mKx-k_0 n_0z]} 
\end{equation}
where $A_m$ is the amplitude of the $m_{th}$ diffracted mode, and  $K=2\pi/a$.  
 
 To analytically solve Eq.~(\ref{equation 2}), in addition to $E(x,z)$ we need to find $n(x)$. Through Eq.~(\ref{equation 1}), supposing $n$ and $\gamma$ are significantly smaller than $n_0$, the RI modulation can be rephrased as:

\begin{equation}
    n(x)=\sqrt{n_0^2+n_0[c^+e^{iKx}+c^-e^{-iKx}}]
    \label{equation 9}
\end{equation}
where $c^+=n+\gamma$ and $c^-=n-\gamma$. Now, by substituting 
 Eq.~(\ref{equation 8}) and Eq.~(\ref{equation 9}) into equation Eq.~(\ref{equation 2}), and considering the Raman-Nath approximation the wave coupled equation (Eq.~(\ref{equation 10})) is obtained \cite{berry1998diffraction, gao2018intrinsic, berry1998lop}.

 \begin{equation}
    \frac{\partial A_m}{\partial z}+\frac{im^2K^2A_m}{2k_0n_0}-\frac{ik_0}{2}[c^-A_{m+1}+c^+A_{m-1}]=0
    \label{equation 10}
 \end{equation}
 
 In contrast with Bragg gratings, in which only two first orders ($m=0,1$) at Bragg angle show non-zero amplitude, higher orders can indicate non-zero amplitude in thin gratings. However, in this active PT grating work with a normal incident light input, the amplitude of higher orders ($m=\pm 2$) are set to be significantly small \cite{berry1998diffraction}. Therefore, neglecting higher orders, which give us a reasonable approximation of the exact solution and provide a clear insight into the PT-symmetry concept \cite{gao2018intrinsic}.   
 Hence, Eq.~(\ref{equation 10}) reduced to a set of coupled equations  (Eq.~(\ref{equation 11})), which can also be written in a matrix form with Hamiltonian $H$, shown in Eq.~(\ref{equation 12}).

\begin{equation}
\label{equation 11}
\begin{split}
 \frac{\partial A_0}{\partial z}-\frac{ik_0}{2}[c^-A_{+1}+c^+A_{-1}]=0 \\ 
\frac{\partial A_{+1}}{\partial z}+\frac{iK^2A_{
+1}}{2k_0n_0}-\frac{ic^+k_0A_0}{2}=0\\
\frac{\partial A_{-1}}{\partial z}+\frac{iK^2A_{-1}}{2k_0n_0}-\frac{ic^-k_0A_0}{2}=0
\end{split}
\end{equation}

\begin{equation}
\label{equation 12}
H=
\begin{bmatrix}
\sigma & -k_0c^+/2 & 0\\
-k_0c^-/2 & 0 &  -k_0c^+/2\\
0 & -k_0c^-/2 & \sigma
\end{bmatrix}
\end{equation}
 where $\sigma=K^2/2k_0n_0$. The eigenvalues of this Hamiltonian, or in other words, the amplitude of the diffracted modes, are:

 \begin{gather}
\lambda_{1}=\sigma \\ 
\lambda_{2}=1/2[\sigma-\sqrt{4k_0^2(n^2-\gamma^2)+\sigma^2}]\\
\lambda_{3}=1/2[\sigma+\sqrt{4k_0^2(n^2-\gamma^2)+\sigma^2}]
\end{gather}

While $\lambda_1$ is independent of the real and imaginary parts of the refractive index modulation, $\lambda_2$ and $\lambda_3$ are dependent on the real and imaginary parts. Thus, once Eq.~(\ref{equation 16}) is satisfied, the system will be located at its exceptional point, and $\lambda_2$ and $\lambda_3$ and their related eigenstates coalesce. 

\begin{equation}
\gamma=\sqrt{n^2+\frac{\sigma^2}{4k_0^2}}
\label{equation 16}
\end{equation}

\begin{figure}[t]
  \centering
   {\includegraphics[scale=0.44]{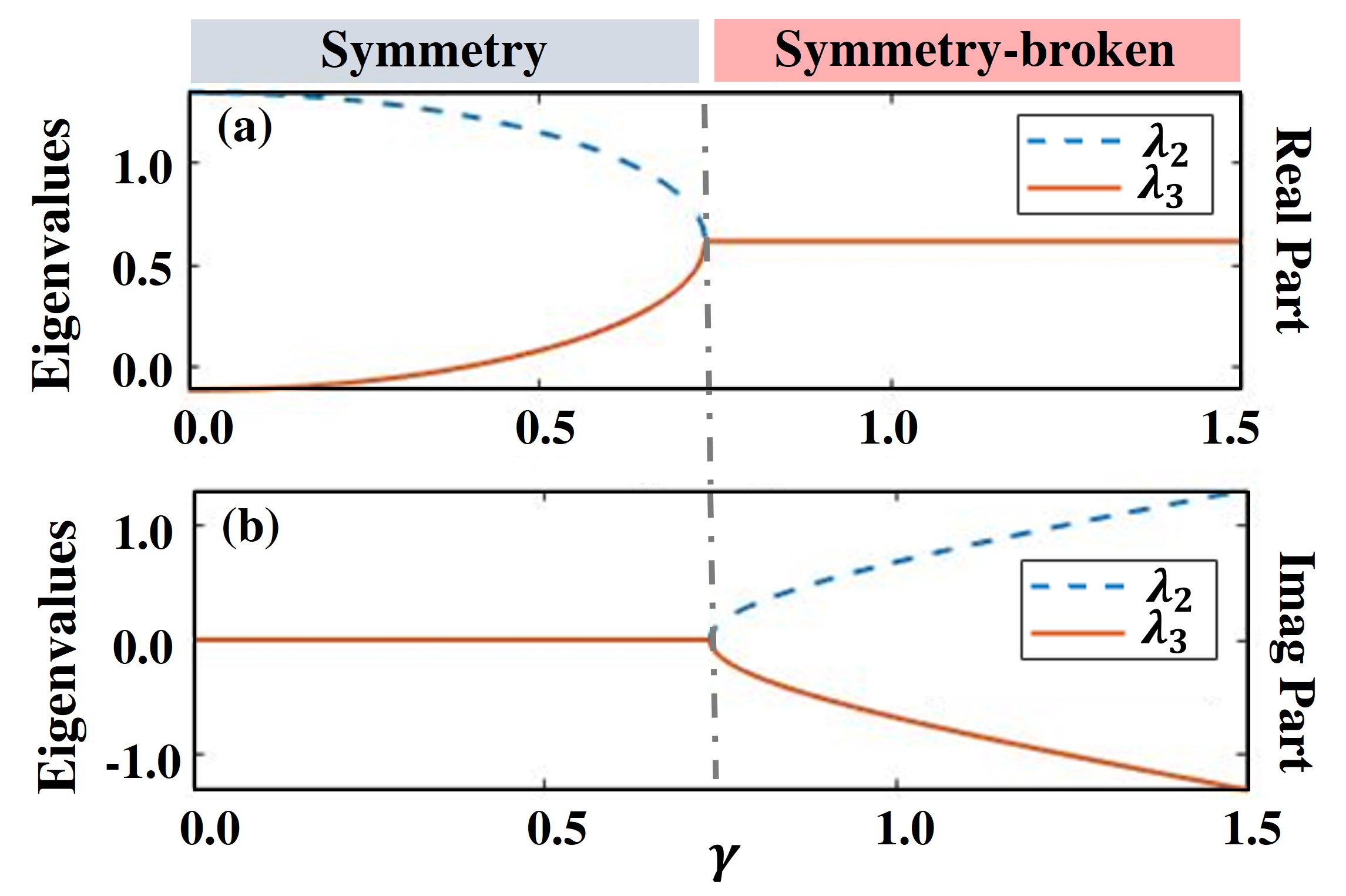}}
  \caption{ (a) The real part and, (b) The imaginary part of eigenvalues according to the $\gamma$ for a system with the periodicity of 2 $\mu m$. The exceptional point is shown by the dashed gray line.}
  \label{fig:figure 2}
\end{figure}

Therefore, once the imaginary part of  the refractive index modulation is $\mid \gamma \mid$ $>$ $\sqrt{n^2+\frac{\sigma^2}{4k_0^2}}$, the system enters into the symmetry-broken phase, indicating imaginary eigenvalues. Fig. \ref{fig:figure 2}(a) and Fig. \ref{fig:figure 2}(b) show the real and imaginary parts of eigenvalues according to the imaginary part of the refractive index ($\gamma$), where the dashed blue line indicates $\lambda_2$ and the solid red line displays $\lambda_3$. As we can see, in the symmetry region, both eigenvalues show a pure real value. However, immediately after the exceptional point, shown here by the dashed gray line, two eigenvalues bifurcates so that one of them possesses a negative imaginary part ($\lambda_3$) and the other one indicates a positive imaginary part ($\lambda_2$). It can be interpreted that one mode is trapped mostly in gain and the other one captured in loss and suppressed. The bifurcation is the most remarkable property of PT-symmetric systems \cite{dong2019symmetry, yang2014symmetry}. This unique property is used in designing single-mode lasers \cite{zhu2019laser, feng2014single}, where undesired competing modes are captured in the loss and suppressed, and only confined modes in the gain are amplified.

\begin{figure}[h]
  \centering
   {\includegraphics[scale=0.44]{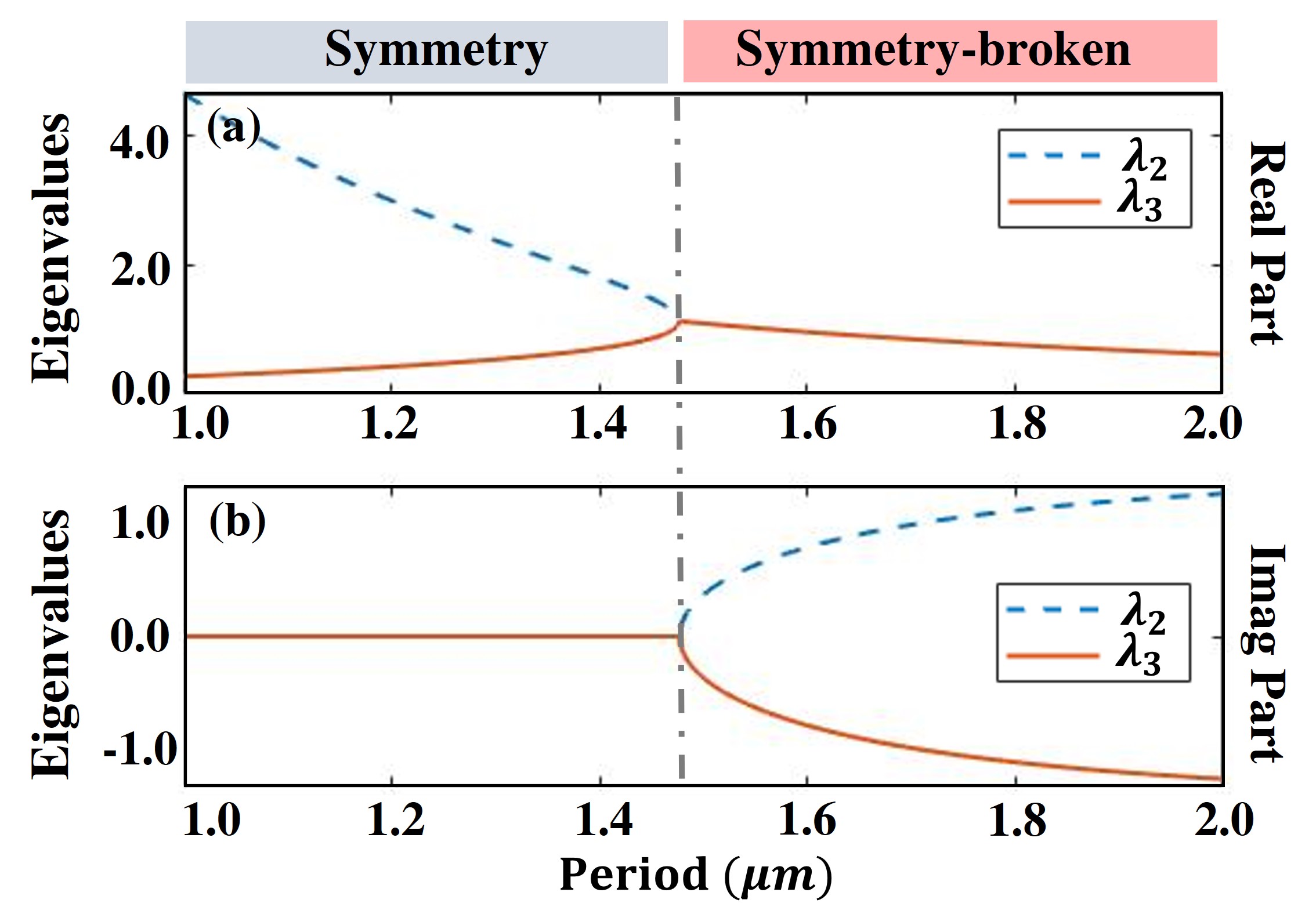}}
  \caption{ (a) The real and, (b) The imaginary parts of eigenvalues according to the period for a PT-symmetric grating. The exceptional point is shown by the dashed gray line.}
  \label{fig:figure 3}
\end{figure}

In gratings that incident light is parallel to the periodicity (along x-direction in Fig. \ref{fig:figure 1}(b)), only the ratio between real and imaginary parts of  the refractive index ($n$ and $\gamma$) determines the phase of the system \cite{lin2011unidirectional}. However, in our designed grating, where the incident light is perpendicular to periodicity, the period plays an essential role.

Figs. \ref{fig:figure 3}(a) and \ref{fig:figure 3}(b) show the real and imaginary parts of eigenvalues according to different period selections, respectively. This figure indicates that increasing the period will drive the system into the symmetry-broken phase. The dashed gray line shows the location of the exceptional point. As reflected from the Eq.~(\ref{equation 16}) in addition to real and imaginary parts of RI modulation, period and the effective refractive index can alter the system phase from symmetry to symmetry-broken phase. In other words, more design freedom exists in this unique active photonic grating platform.

This section analytically solved the Helmholtz equation for the dominant grating modes considering the Raman-Nath approximation. We showed the impact of the imaginary part of RI modulation and period to shift the system from a PT-symmetry phase to the symmetry-broken stage. In the following section, we will design a thin grating that follows the theory described above, and report the unique PT-symmetric photonic modes solved numerically by the RCWA method. Moreover, the related electric field and amplitude distribution over the grating will be displayed, which will extend the understanding of physical mechanisms leading to extraordinary phenomena observed in the active PT-symmetric photonic grating structures.

\section{\label{sec:level3}Numerical simulations}

\begin{figure}[ht]
  \centering
   {\includegraphics[scale=0.49]{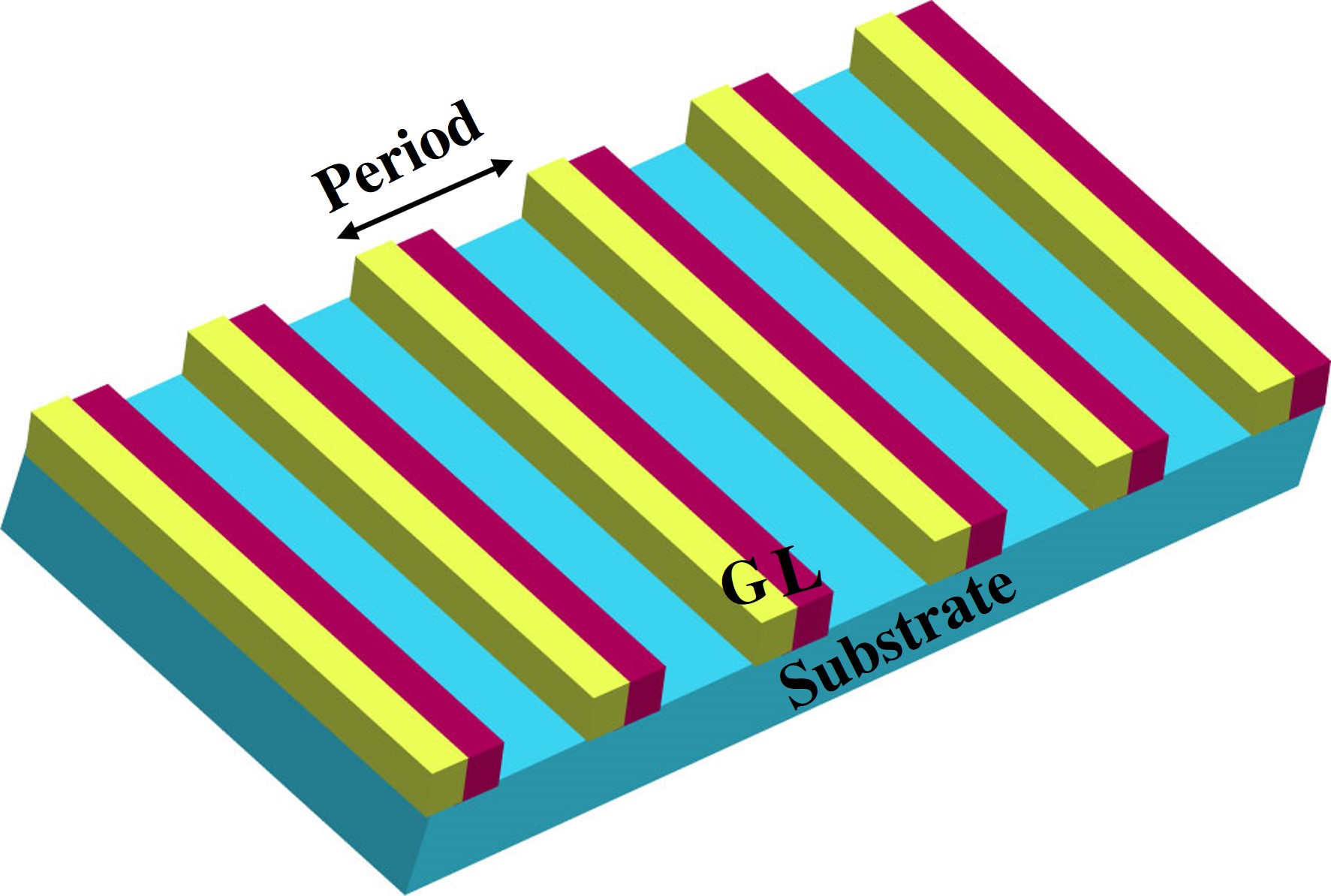}}
  \caption{ A 3D schematic view of the PT-symmetric diffraction gratings.  }
  \label{fig:figure 4}
\end{figure}

Fig. \ref{fig:figure 4} displays a 3D schematic view of the designed diffraction grating, where yellow and red parts show the gain and loss ($\mid$$\gamma$$\mid$), respectively. In this numeral study, we adopt common material parameters from the III-V semiconductors, InGaAs just as an example, to design and study the PT-symmetric grating effects. It is noted that conclusions to be made can also be applied to other materials systems in other spectral domains.

One of the most well-known criteria of PT-symmetric systems is the spatial bifurcation of two modes in the symmetry-broken region. As presented in Fig. \ref{fig:figure 2}, in the symmetry-broken phase, two modes with the same amplitude (same real part of eigenvalues) show the bifurcation, where one is captured mostly in the gain other trapped mainly in the loss. Thus, one mode is amplified while the other one is suppressed. This phenomenon is the primary mechanism behind the single-mode lasing in most of the PT-symmetric lasers. We expect that the numerical simulation demonstrates the same mechanism.

\begin{figure}[b]
  \centering
   {\includegraphics[scale=0.41]{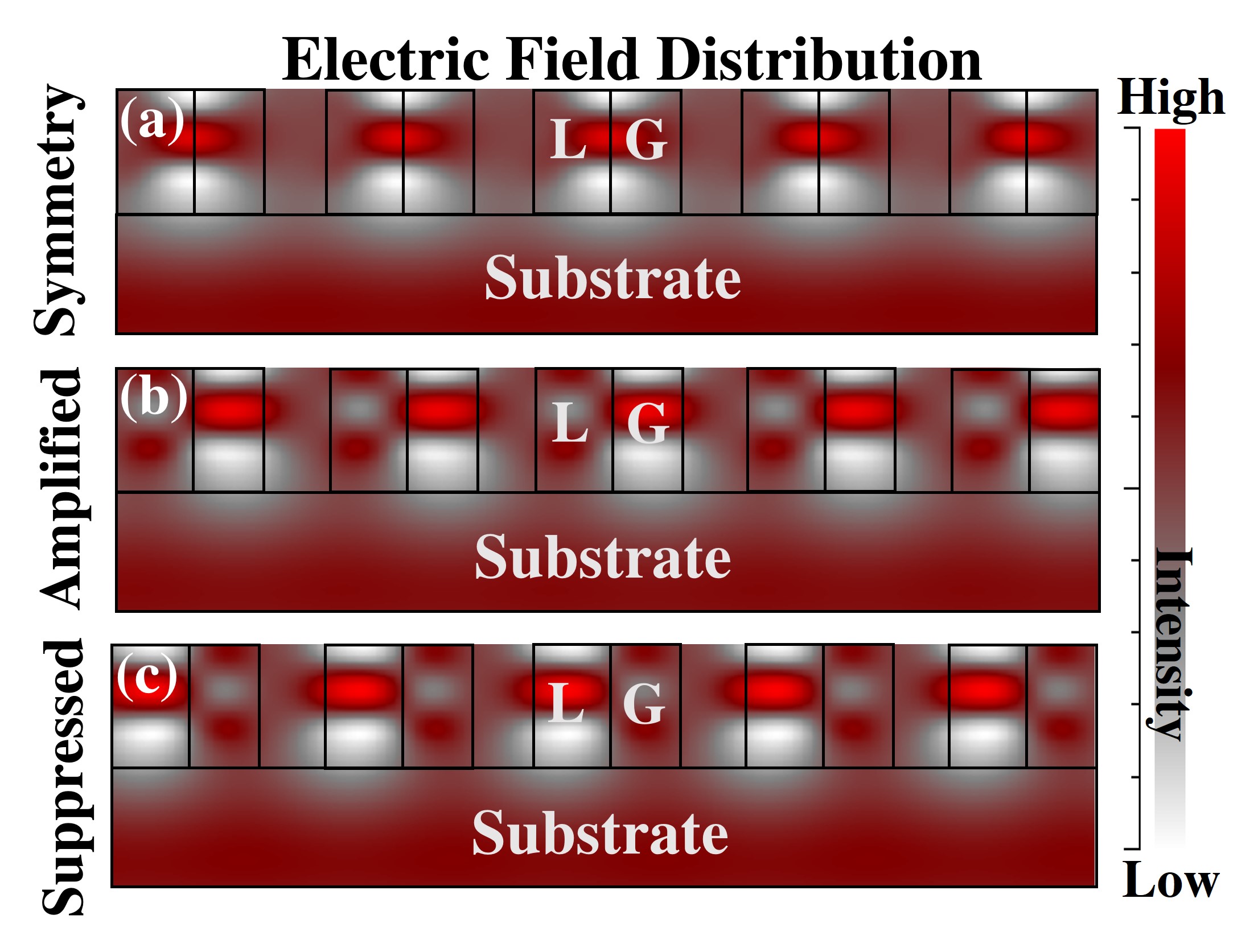}}
  \caption{ The electric field distribution over the PT-symmetric grating where (a) $\gamma=0.2$, associated with the symmetry phase. (b) $\gamma=0.22$ associated with the symmetry-broken phase (amplified mode). (c) $\gamma=0.22$ associated with the symmetry-broken phase (suppressed mode).} 
  \label{fig:figure 5}
\end{figure}

Fig. \ref{fig:figure 5} shows the spatial electric field distribution over a grating with the period of 1 $\mu m$ and the thickness of 821 nm. The substrate and gain/loss areas are demonstrated in this figure, where G stands for gain and L stands for loss.
The grating parameters were selected in a way that satisfies the Raman-Nath approximation criteria (Eq.~(\ref{equation 17})) that was implemented in the theoretical section.

\begin{equation}
\label{equation 17}
    \begin{cases}
     Q^{'}\zeta < 1 \\ 
Q^{'}=\frac{2\pi\lambda d}{n_0 \Lambda^2 cos\theta} \\
\zeta=\frac{\pi n d}{2 \lambda n_0}
    \end{cases}       
\end{equation}

In Eq.~(\ref{equation 17}), $\lambda$ is the free-space wavelength, $d$ is the grating thickness, $n_0$ is the effective refractive index, $\Lambda$ is the grating period, $n$ is the real part of the RI modulation, and $\theta$ is the incident angle.

 Fig. \ref{fig:figure 5}(a) shows the symmetric distribution of the electric field over the gain and loss areas so that no mode experiences net amplification or suppression. This pattern can be associated with the PT-symmetry phase, where both eigenvalues are purely real. By increasing gain/loss ($\gamma$) from 0.2 to 0.22, the system is shifted to the symmetry-broken phase, where one mode is captured in the gain and experiences the amplification. Fig. \ref{fig:figure 5}(b) displays this mechanism clearly. However, the other mode is trapped in the loss and is suppressed (Fig. \ref{fig:figure 5}(c)). 

\begin{figure}[t]
  \centering
   {\includegraphics[scale=0.33]{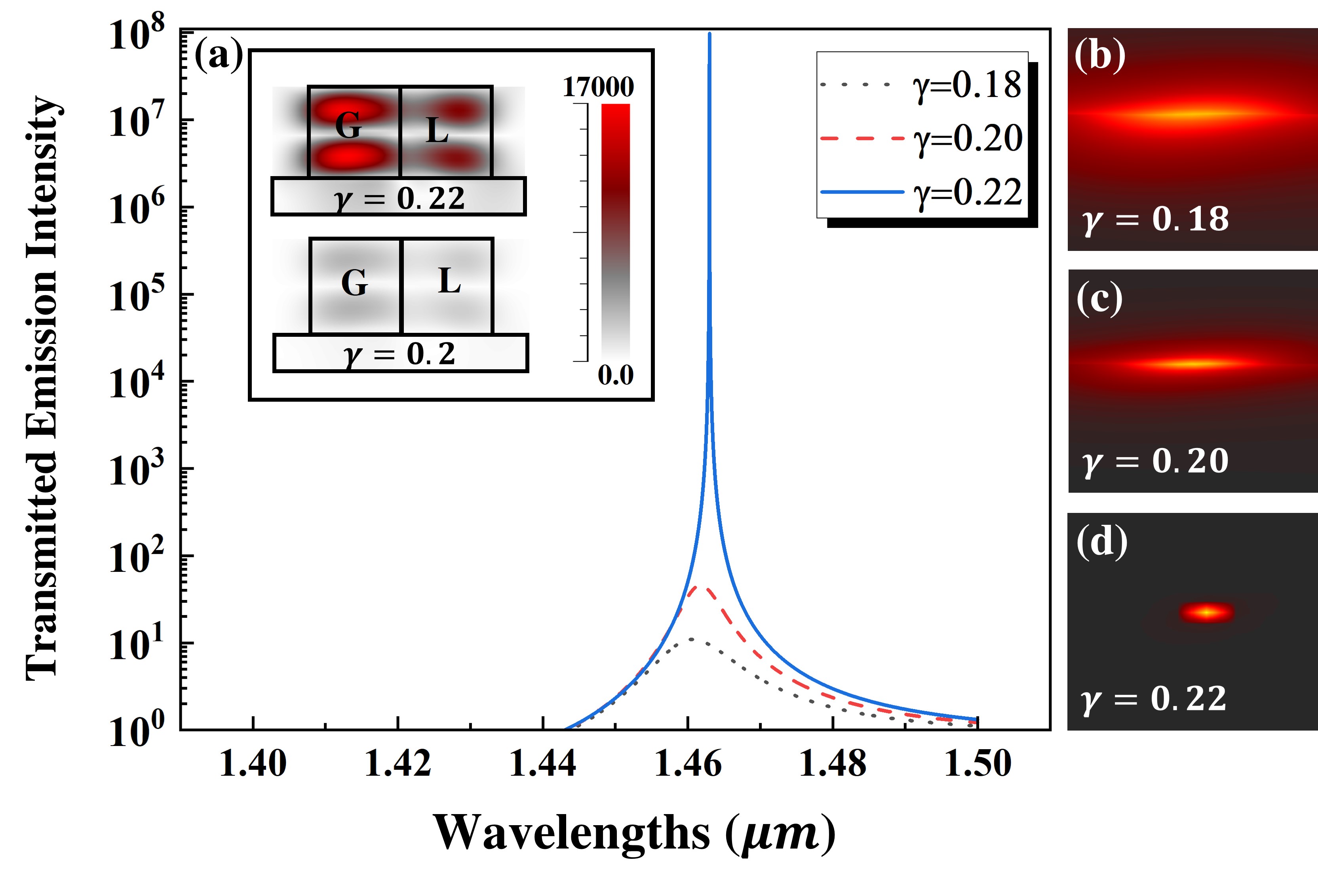}}
  \caption{ (a) The logarithm form of the transmitted spectra over wavelengths from the grating with $\gamma=0.18$ (the black dotted line), $\gamma=0.2$ (the red dashed line), and $\gamma=0.22$ (the solid blue line). The amplitude distribution of the spectrum with $\gamma=0.2$ and $\gamma=0.22$ is displayed over one grating period in inset. The transmitted emission coefficient map over wavelengths where $\gamma$ changes from 0 to (b) 0.18, (c) 0.2, and (d) 0.22. }
  \label{fig:figure 6}
\end{figure}

After indicating the bifurcation over the PT-symmetric diffraction grating, we found out that the system can show a single-mode, almost zero-bandwidth (0.04 nm) mode through optimizing gain/loss contrast ($\gamma$). 
To better understand this non-trivial mode, we compared the diffraction efficiency of the transmitted mode in the logarithm form, where $\gamma=0.18$, $\gamma=0.2$, and $\gamma=0.22$. Fig. \ref{fig:figure 6}(a) illustrates this comparison, where the solid blue line is related to the system with $\gamma=0.22$, the dashed red line is associated with $\gamma=0.2$, and the black dotted line is associated with $\gamma=0.18$. The inset shows field amplitude distribution over a unit cell in a grating with $\gamma=0.2$ and $\gamma=0.22$. The magnitude of amplitude is displayed by the colored bar. Furthermore, Fig. \ref{fig:figure 6}(b), Fig. \ref{fig:figure 6}(c), and Fig. \ref{fig:figure 6}(d) show the 2D transmitted emission coefficient map over wavelengths, where $\gamma$ changes from 0 to 0.18, 0.2, and 0.22, respectively. As we can see, the broadness of the mode decreases by increasing $\gamma$, which is matched with the full-width half maximum of the spectrum displayed in Fig. \ref{fig:figure 6}(a)

As Fig. \ref{fig:figure 6}(a) shows, the transmitted mode when $\gamma=0.22$ is $10^6$ greater than when $\gamma=0.2$.  The primary reason for this huge difference can be related to the intensity of the confined field in the gain area. The inset illustrates that the amplitude of the confined mode in the gain area for the grating with $\gamma=0.22$ is almost $10^3$ greater than the amplitude of the trapped mode in the gain for the same grating but with $\gamma=0.2$. The physical mechanism for this extraordinary phenomenon has roots in the scattering theory in PT-symmetric systems and the concept of spectral singularity \cite{mostafazadeh2015physics,chaos2013resonant,mostafazadeh2009spectral}, which will be discussed in detail in the following paragraphs.

The properties of gratings as scattering systems can be defined by the Scattering matrix (S-matrix). This matrix relates the input field to the output field, passing through a scattering system. Eq.~(\ref{equation 18}) shows the S-matrix ($S_{nm}$) that associates the incoming mode amplitude ($\phi_{m}$) to the outgoing mode amplitude ($\eta_{n}$). 

\begin{equation} \label{equation 18}
{\sum_{n}\ S_{nm} (\omega) \phi_{m}= \eta_{n}}
\end{equation}
The required condition to define an S-matrix is the unitary condition ($\mid$det S$\mid$=1). For passive structures with real frequencies, this unitary condition is only satisfied by having unimodular S-matrix eigenvalues ($\mid$$s_\pm$$\mid$=1) \cite{linton2009wave}. Applying PT operator on S-matrix meets the unitary condition. However, in contrast with passive structures, the only way to meet the unitary condition in PT-symmetric systems is not having unimodular eigenvalues. In these systems, the unitary condition can be satisfied by pairs reciprocal moduli as well ($s_\pm$=$1/s^*_\mp$) \cite{chong2011p}. The former possibility to satisfy the unitary condition is related to the PT-symmetry (unimodular eigenvalues), and the latter way (reciprocal eigenvalues) is associated with the PT-symmetry-broken.

In the PT-symmetry-broken phase, there are points at which poles and zeros of the S-matrix intersect on the real axis \cite{chong2010coherent, baranov2017coherent}. This intersection leads to approaching $s_n$ to zero and consequently $1/s^*_n=\infty$, while their product is still unity. In fact, these singular points, named spectral singularity, correspond to the giant amplitude enhancement of the transmitted emission modes for an optimized value of gain/loss.
\begin{figure}[b]
  \centering
   {\includegraphics[scale=0.35]{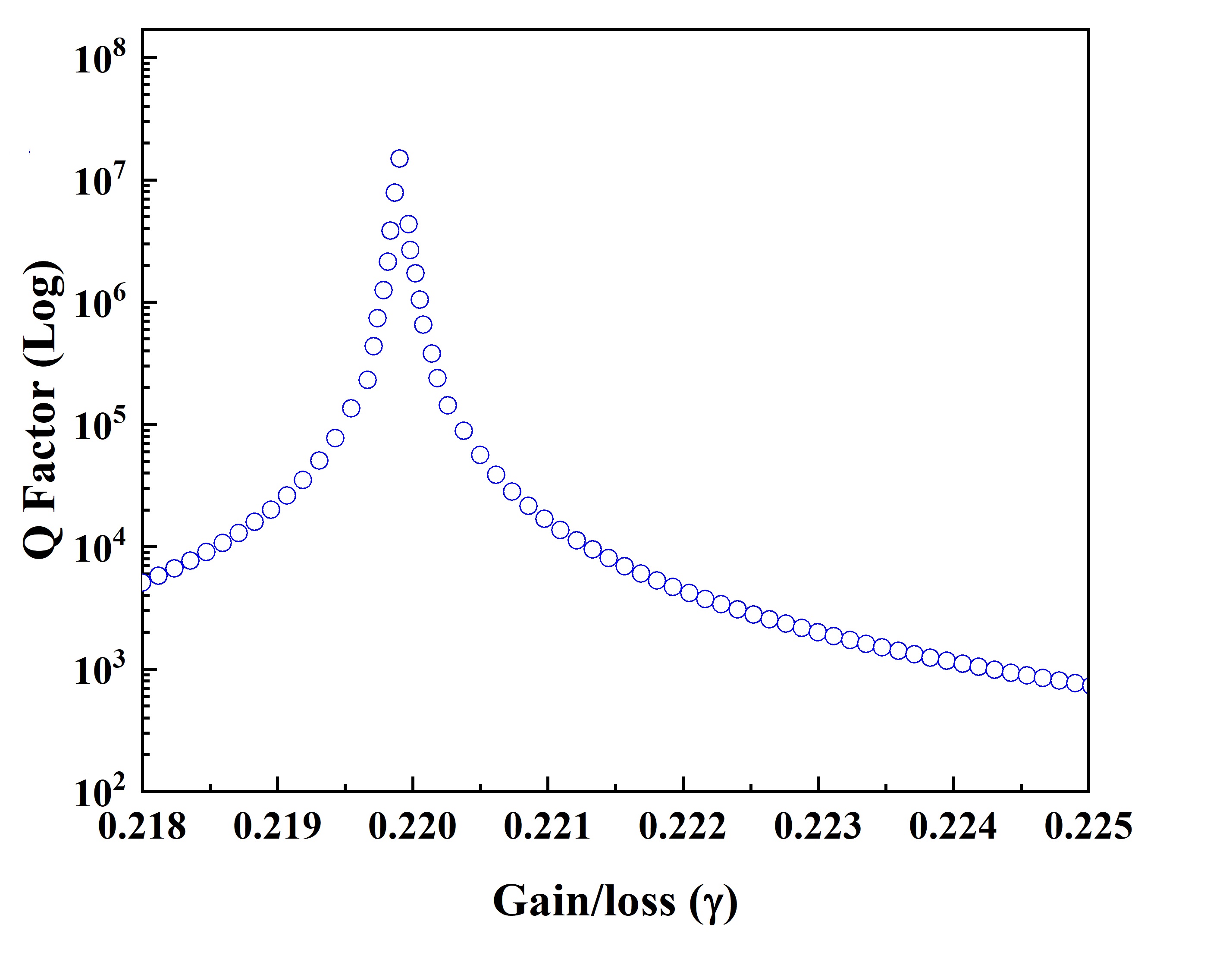}}
  \caption{The Q factor of the resonance mode in the logarithm form with  different  introduced  gain/loss ($\gamma$) to the  grating.}
  \label{fig:figure 7}
\end{figure}
SSs are a family member of resonant modes \cite{mostafazadeh2009resonance, mostafazadeh2011optical}. However, the main feature for SSs that distinguishes them from trivial resonances is eliminating the imaginary part of the resonant frequencies \cite{mostafazadeh2015physics}. The imaginary part of the resonant frequencies is responsible for evanescent energy or, in other words, the bandwidth of the resonant peaks, the lower evanescent energy, narrower bandwidth. The uniqueness of SSs lies in being purely real, which in theory leads to zero-bandwidth resonances \cite{mostafazadeh2015physics}. It should be noted that SSs are different from the Bound States In Continuum (BICs); however, both show zero-bandwidth resonances \cite{mostafazadeh2012spectral}.

Fig. \ref{fig:figure 7} indicates the Q factor of the resonant mode (1.463 $\mu m$) with different introduced gain/loss ($\gamma$) to the grating. A sharp enhancement is observed when $\gamma$ touches 0.22. 
This figure shows that only at a specific gain/loss value, here $\gamma=0.22$, SS can occur, and higher or lower values cannot satisfy the SS condition. This is matched with the theory that claims only and only at the SS point, $s_n$ goes to zero, while $1/s^*_n$ goes to $\infty$, and the giant transmitted emission enhancement is related to $1/s^*_n$, which its amplitude approaches infinity. This transmitted emission coefficient map indicates that by this unique active PT grating design, $\gamma=0.22$ leads to appearing one SS at 1.46 $\mu m$.

\section{\label{sec:level5}Conclusion}
In this work, we report a new active granting platform having the optical gain and loss to realize non-trivial PT-symmetry modulation effects. Herein, by solving the Helmholtz equation in the Raman-Nath approximation, the bifurcation leading to confining the amplified mode in the gain region was indicated. We showed that in the PT photonic grating structure, in addition to the gain/loss parameter, the period selection can also drive the system from PT-symmetry to the symmetry-broken phase. Moreover, a zero-bandwidth photonic emission mode (i.e., at 1.463 $\mu m$) can be stimulated and numerically demonstrated through the RCWA simulation. It is suggested that, by optimizing the optical gain/loss contrast, this non-trivial high Q-factor resonance mode is the result of the photonic system's special spectral singularity effect, where one of the scattering states approaches zero while its reciprocal state approaches infinity. Considering the functionality of gratings in a broad range of applications, we anticipate introducing active PT-symmetric gratings will offer new engineering strategies for developing new low-threshold and super-coherent surface-emitting lasers, and thus, advancing modern on-chip photonic applications.

\section{ACKNOWLEDGMENTS}
The authors would like to thank Dr. Mykola Kulishov for the insightful discussions. B.W. acknowledges the partial supports from OU's Big Idea Challenge award and the Oklahoma Center for the Advancement of Science and Technology's Research Grant \#AR21-052. Y. Z. acknowledges the partial supports from the Natural Science Foundation of Shanghai (21ZR1443100) and the National Natural Science Foundation of China (61705099).

\bibliography{ref}

\end{document}